\documentclass{article}
\usepackage[utf8]{inputenc}
\usepackage{amsfonts}
\usepackage{spconf,amsmath,graphicx}
\usepackage{xcolor}
\graphicspath{{./}}
\usepackage{subfigure}
\usepackage[T1]{fontenc}
\usepackage{csquotes}
\usepackage{float}
\usepackage{bm}

\name{Qing Zou$^*$, Abdul Haseeb Ahmed$\dag$, Prashant Nagpal$^\dagger$, Sarv Priya$^*$, Rolf F Schulte$^{**}$, Mathews Jacob$^*$.}
\address{$^*$University of Iowa, $^\dag$ Philips HealthCare, $^\dagger$ University of Wisconsin, $^{**}$ GE HealthCare}
\begin{document}

\title{Joint alignment and reconstruction of multislice dynamic MRI using variational manifold learning}

\maketitle

\begin{abstract}
	Free-breathing cardiac MRI schemes are emerging as competitive alternatives to breath-held cine MRI protocols, enabling applicability to pediatric and other population groups that cannot hold their breath. Because the data from the slices are acquired sequentially, the cardiac/respiratory motion patterns may be different for each slice; current free-breathing approaches perform independent recovery of each slice. In addition to not being able to exploit the inter-slice redundancies, manual intervention or sophisticated post-processing methods are needed to align the images post-recovery for quantification. To overcome these challenges, we propose an unsupervised variational deep manifold learning scheme for  the joint alignment and reconstruction of multislice dynamic MRI. The proposed scheme jointly learns the parameters of the deep network as well as the latent vectors for each slice, which capture the motion-induced dynamic variations, from the k-t space data of the specific subject. The variational framework minimizes the non-uniqueness in the representation, thus offering improved alignment and reconstructions. 
	
\end{abstract}

\section{Introduction}

Breath-held cine MRI is an integral part of clinical cardiac exams because it can offer valuable indications of abnormal structure and function of the heart. However, this approach is often infeasible for several subject groups that cannot hold their breath, including pediatric and chronic obstructive pulmonary disease (COPD) subjects. Several free-breathing imaging schemes, which rely on navigated or self-gating \cite{rosenzweig2020cardiac,Zhou2019}, kernel manifold approaches \cite{nakarmi2017kernel,ahmed2020free}, and deep learning based manifold models \cite{zou2021dynamic,zou2021deep,yoo2021time} were introduced to overcome the above challenge. The navigated or self-gating methods derive gating signals from the measured data, which is used to bin the acquired data to different cardiac or respiratory phases; the binned data is used to recover the images in different cardiac and respiratory phases. Manifold approaches follow a conceptually similar approach, except that the data is not explicitly binned; the manifold structure of images is exploited using a kernel-based formulation \cite{nakarmi2017kernel,ahmed2020free}. Recently, unsupervised deep generative models that exploit the manifold structure of images were shown to outperform the classical manifold methods  \cite{zou2021dynamic,zou2021deep,yoo2021time}. This scheme relies on a deep convolutional neural network (CNN) that generates the 2D images when excited with low-dimensional latent vectors that capture the intrinsic variability in the dataset, including cardiac and respiratory motion. The patient-specific CNN parameters and the time series of latent vectors are learned from the k-t space data of the subject. The loss is the mean square error between the k-t space measurements of the generated image series and the k-t space measurements. Once learned, the generator is excited with the learned latent vectors when it generates the image time series.  

 All of the above-mentioned navigated and manifold approaches consider the independent recovery of the data from each slice. Cardiac MRI data is often acquired in the multi-slice mode rather than as a 3D acquisition to achieve good in-plane and temporal resolution, as well as to preserve good  blood-pool to myocardium contrast resulting from inflow effects. Since the slices are acquired sequentially, the cardiac and respiratory motion patterns are often drastically different between the slices. A challenge with the above independent recovery strategy is the inability of these schemes to exploit the inter-slice redundancies for image recovery. In addition, manual post-processing may be needed to align the data from different phases and slices. 

The main focus of this work is to extend the 2D deep generative models \cite{zou2021dynamic,zou2021deep} for the joint alignment and recovery of 3D multislice data. In particular, we consider a 3D CNN that generates a 3D image volume when driven by a low-dimensional latent vector. We propose to learn the generator parameters and the latent vector time series corresponding to each slice jointly from the multislice k-t space data from all the slices. We expect the latent vectors of each slice to capture the cardiac and respiratory motion patterns during the acquisition of that specific slice, while the same generator is used for all slices (see Fig. \ref{illu} for an illustration). Once learned, we will excite the generator with latent vectors of any slice when we expect the generator to recover aligned image time series. A challenge with this direct extension of \cite{zou2021dynamic,zou2021deep} to the 3D setting is the non-uniqueness of the learned representation and the latent vectors. For instance, one may scale the latent vectors and correspondingly modify the CNN weights to obtain the same representation. Without additional constraints, the learned latent vectors corresponding to different slices will follow different probability distributions; driving the 3D generators with any of the latent vector sets will only guarantee good reconstructions for that specific slice and often results in poor reconstructions for other slices. 

We introduce a novel variational scheme to learn the deep 3D generative model directly from the single-slice measurements to minimize the above issues resulting from the non-uniqueness of the representation. The variational approach allows us to impose priors on the latent vectors, which will encourage the latent vectors of different slices to have the same distribution. This approach can be viewed as the generalization of the variational auto-encoder (VAE) to the undersampled setting. In particular, we estimate the latent variables and the CNN weights using a maximum likelihood formulation. Since the likelihood is not tractable, we maximize its variational lower bound involving a zero mean unit Gaussian probability distribution for the latent variables. The classical VAE scheme uses an encoder network to derive the conditional probabilities of the latent vectors from fully sampled images. This approach is not directly applicable in our setting without fully sampled images; each image is measured using a different measurement operator. We instead model the conditional density as a Gaussian distribution whose parameters are learned from the undersampled data directly using back-propagation. We use a Gaussian prior on the latent variables while deriving the evidence-based lower bound (ELBO); the Gaussian prior ensures that the latent variables from different slices have similar distributions and facilitates the alignment of the slices.

\section{Proposed Approach}

\begin{figure}[!t]
\centering
          % \subfigure[V-SToRM: SS]{\includegraphics[width=0.3\textwidth]{figs/illu_vSToRM_ss}}\hspace{5em}
	%\subfigure[V-SToRM: MS]{\includegraphics[width=0.36\textwidth]{figs/illu_vSToRM_ms}}\vspace{-1.5em}
	\includegraphics[width=0.5\textwidth]{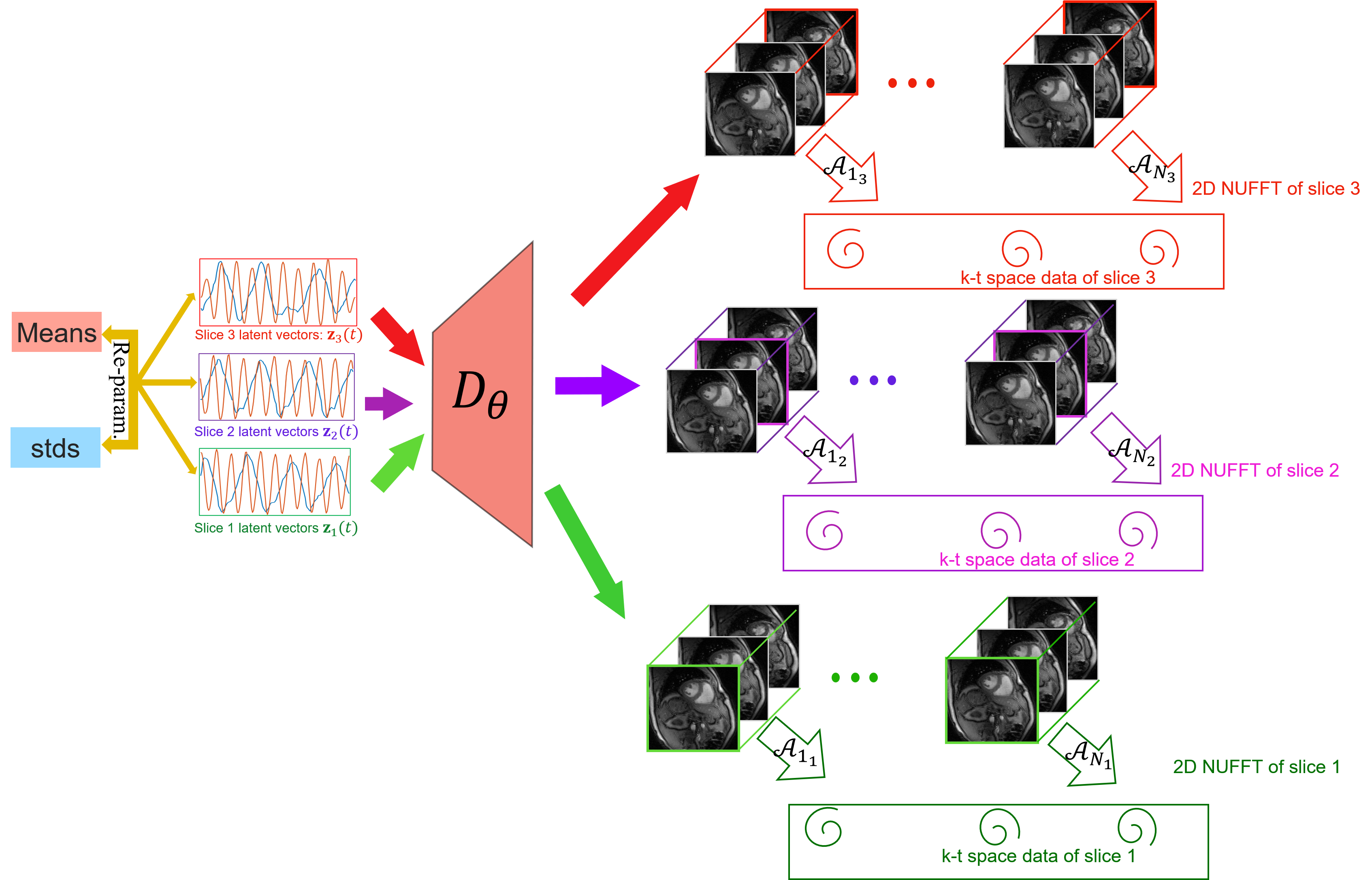}\vspace{-1em}
	\caption{\small{Illustration of the proposed variational SToRM (V-SToRM) scheme. The inputs to the 3D network are samples from the respective latent distributions. The 3D volumes are sampled by the respective sampling operators $\mathcal A_{z,t}$, which extract the $z^{\rm th}$ slice and compare to the measured data. The optimization criterion specified by \eqref{multislice} is minimized in this case.}}
	\label{illu}
\end{figure}

We consider the reconstruction and alignment of  the 3D cardiac volumes of the heart from its undersampled 2D multislice k-t space data in this work. Specifically, we consider the recovery of the time series $\mathbf{x}(\mathbf{r},t_z)$, where $\mathbf r=(x,y,z)$ represents the spatial coordinates and $t_z$ denotes the  timeframe during the acquisition of the $z^{\rm th}$ slice . We model the acquisition of the data as 
\begin{equation}
\mathbf b(t_z) = \mathcal{A}_{t_z}\Big(\mathbf{x}(\mathbf r,t_z)\Big) + \mathbf n_{t_z},
\end{equation}
where $\mathbf b(t_z) $ is the k-t space data of the $z^{\rm th}$ slice at the $t^{\rm th}$ timeframe and $\mathcal{A}_{t_z}$ are the time-dependent measurement operators, which evaluate the multi-channel single-slice Fourier measurements of the 3D volume $\mathbf x(\mathbf r,t_z)$ on the trajectory $k_{t_z}$ corresponding to the time point $t$. $\mathbf n_{t_z}$ represents the noise in the measurements. We model the volumes in the time series as
\begin{equation}\label{key}
\mathbf x_i = \mathcal D_{\theta}(\mathbf c_i),
\end{equation}
where $\mathbf c_i$ is latent variable corresponding to $\mathbf x_i$ and $\mathcal D_{\theta}$ is a CNN-based generator with parameters $\theta$, which are shared for all image volumes.

We consider the recovery of the images $\mathbf x_i$ from their measurements by maximizing the likelihood:
\begin{equation}\label{likelihood}
p(\mathbf{b}_i|\theta)= \frac{p(\mathbf{b}_i,\mathbf{c}_i|\theta)}{p(\mathbf{c}_i|\mathbf{b}_i,\theta)}
\end{equation}
We note that the posterior $p(\mathbf{c}_i|\mathbf{b}_i)$ is not tractable. Following the VAE approach in \cite{kingma2013auto}, we use a surrogate distribution to approximate $p(\mathbf{c}_i|\mathbf{b}_i)$. We propose to use a Gaussian model $q_i(\mathbf{c}_i) \approx p(\mathbf{c}_i|\mathbf{b}_i,,\theta)$, parameterized by its mean $\boldsymbol\mu_i$ and diagonal covariance matrix $\boldsymbol\Sigma_i$, to approximate the posterior $p(\mathbf{c}_i|\mathbf{b}_i)$. The mean and the variance will then be learned using back-propagation. Following a similar argument as in \cite{kingma2013auto}, the likelihood term in \eqref{likelihood} can be lower bounded as 
\begin{eqnarray}\label{elbo}\nonumber
\log p(\mathbf{b}_i|\theta) &\ge& \underbrace{-\frac{1}{2\sigma^2}\mathbb{E}_{\mathbf c_i \sim q_i(\mathbf c_i)}\left[ \|\mathcal{A}_i \,\mathcal D_{\theta}(\mathbf{c}_i)-\mathbf b_i\|^2\right]}_{{\text{data term}}} \\
&&\qquad -\qquad \underbrace{KL[q_i(\mathbf c_i)||p(\mathbf{c}_i)]}_{L( q_i):\text{latent regularization}}.
\end{eqnarray}
Here, $p(\mathbf c_i)$ is a prior on the latent variables $p(\mathbf c_i) = \mathcal{N}(\mathbf{0},\mathbf{I})$, where $\mathbf{I}$ is the identity matrix. In this case, the Kullback-Leibler divergence can be explicitly evaluated as
\[L(\mathbf c_i) = \frac{1}{2}\left[-\log[\det(\mathbf{\Sigma})] - n + {\mathrm{trace}}(\mathbf{\Sigma}) + \bm{\mu}^T\bm{\mu}\right],\] 
where we assume a latent space of dimension $n$. We then solve for the unknown weights of the generator $\theta$, $\boldsymbol\mu_i$ and $\boldsymbol\Sigma_i$ by minimizing the negative of the lower bound in \eqref{elbo}. Following \cite{kingma2013auto}, we use the reparameterization trick to sample $\mathbf c_i$ as
\begin{equation}\label{resampling}
\mathbf c_i = \boldsymbol \mu_i + \boldsymbol \Sigma_i ~\boldsymbol\epsilon,
\end{equation}
where $\boldsymbol \epsilon = \mathcal{N}(\mathbf{0},\mathbf{I})$. %The estimation process then involves the minimization of the cost function
%\begin{equation}\label{vloss}
%\mathcal{C}\left(\theta,\{\underbrace{\boldsymbol\mu_i, \boldsymbol \Sigma_i}_{q_i}\}\right) = \sum_{i=1}^{N_{\rm data}} \left(\|\mathcal{A}_i \,\mathcal D_{\theta}(\mathbf{c}_i)-\mathbf b_i\|^2 +  \sigma^2~ L(q_i) \right),
%\end{equation}
%with respect to the unknowns $\theta,\boldsymbol\mu_i$ and $\boldsymbol \Sigma_i$. 
%When we are dealing with single-slice data, the parameters in the generator and the $\boldsymbol\mu_i$ and the $\boldsymbol \Sigma_i$ are updated based on the loss function 
%\begin{equation}\label{singleslice}
%\mathcal{L}(\theta,\{\boldsymbol\mu_t, \boldsymbol \Sigma_t\}) = \mathcal %C(\theta,\{\boldsymbol\mu_t, \boldsymbol \Sigma_t\}) + \lambda_1||\theta||_1^2 + \lambda_2||\nabla\boldsymbol{\mu}_t||^2.
%\end{equation}
%Here, $\mathcal C(\theta,\{\boldsymbol\mu_t, \boldsymbol \Sigma_t\})$ is defined in \eqref{vloss}. 

We propose to jointly align and reconstruct the multislice MRI by jointly estimating the parameters $\theta$, $\boldsymbol{\mu}(t_z)$ and $\boldsymbol \Sigma(t_z)$ from the measured multislice data by minimizing the following loss function:
\begin{align}\label{multislice}\nonumber
\mathcal{L}_{MS}(\theta,\boldsymbol{\mu}(t_z),\boldsymbol \Sigma(t_z)) = & \mathcal C_{MS}(\theta,\boldsymbol{\mu}(t_z),\boldsymbol \Sigma(t_z)) +  \lambda_1||\theta||_1^2\\
& \qquad + \lambda_2\sum_z||\nabla_{t_z}\boldsymbol{\mu}(t_z)||^2,
\end{align}
where
\begin{equation*}
\mathcal C_{MS} = \displaystyle\sum_{z=1}^{N_{\rm slice}}\sum_{t=1}^{N_{\rm data}}\|\mathcal{A}_{t_z}\left[\mathcal D_{\theta}(\mathbf c(t_z))\right] - \mathbf b_{t_z}\|^2+\sigma^2~ L(q(t_z))
\end{equation*}
is the lower bound for maximum likelihood as the first term in \eqref{multislice}. The illustration of this scheme is given in Fig. \ref{illu}. The second term in \eqref{multislice} is a regularization penalty on the generator weights. It has been shown in \cite{zou2021dynamic} that adding this term makes the training of the decoder more stable. The third term involves the temporal gradients of the latent vectors, which enforces the latent vectors to capture the smooth nature of motion patterns in the dynamic images. We use the Adam optimization to determine the optimal parameters. We also adopt the progressive-in-time training strategy introduced in \cite{zou2021dynamic} to realize a computationally efficient reconstruction.

\section{Experiments \& Results}
\vspace{-1em}
\subsection{Datasets and imaging experiments}

The datasets used in this work are acquired using a 2D (GRE) sequence with golden angle spiral readouts in the free-breathing and ungated setting on a GE MR750W scanner. The sequence parameters for the datasets are: FOV = 320 mm $\times$ 320 mm, flip angle = 18$^{\circ}$, slice thickness = 8 mm. A total number of 3,192 spirals were acquired for each slice in the subjects with TR=8.4 ms. Among the 3,192 spirals, every sixth spiral was acquired with the same angle; these spirals were used for self-navigation in the reconstruction methods that require self-navigation. We binned the data from six spiral interleaves corresponding to 50 ms temporal resolution.

%\subsection{Single-slice V-SToRM scheme}

%In this section, we focus on single-slice V-SToRM; the reconstructions of a dataset and its latent vectors are shown in Fig. \ref{ss_showcase}. We trained the variational model using the data of one slice. The latent vectors we obtained are shown at the bottom of Fig. \ref{ss_showcase}. Four different phases in the time series are shown in the figure, and their corresponding latent vectors are indicated in the plot of the latent vectors.

%\begin{figure}[!htpb]
%	\centering
%	\includegraphics[width=0.3\textwidth]{figs/ss_showcase}
%	\caption{\small Showcase of the single-slice V-SToRM.}
%	\label{ss_showcase}\vspace{-1em}
%\end{figure}

\subsection{Multislice V-SToRM scheme}

\begin{figure}[!t]
	\centering
	\subfigure[Reconstructions of slice 3]{\includegraphics[width=0.325\textwidth]{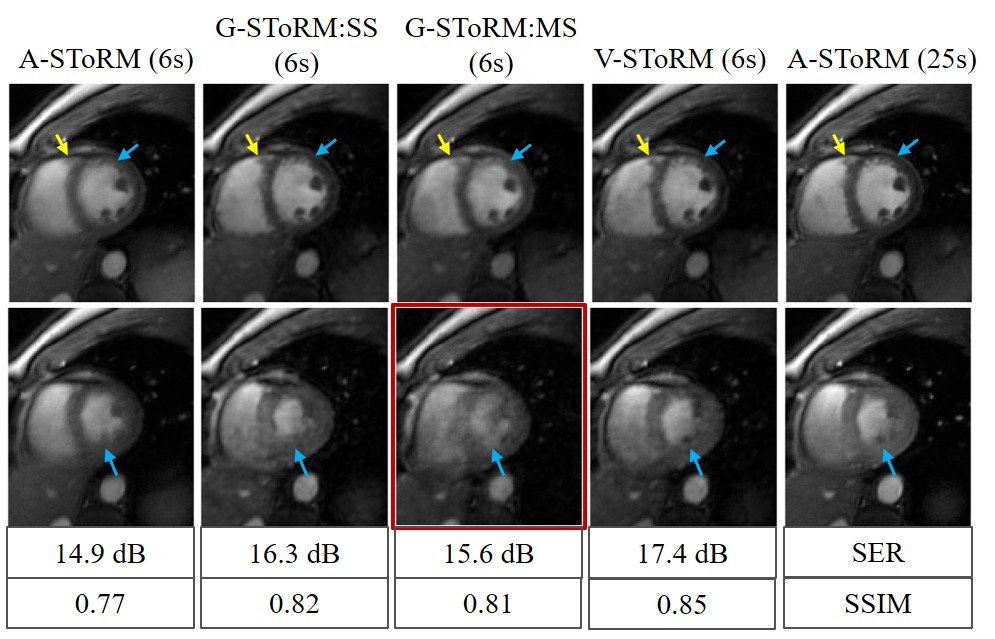}}\\\vspace{-0.7em}
	\subfigure[Reconstructions of slice 4]{\includegraphics[width=0.325\textwidth]{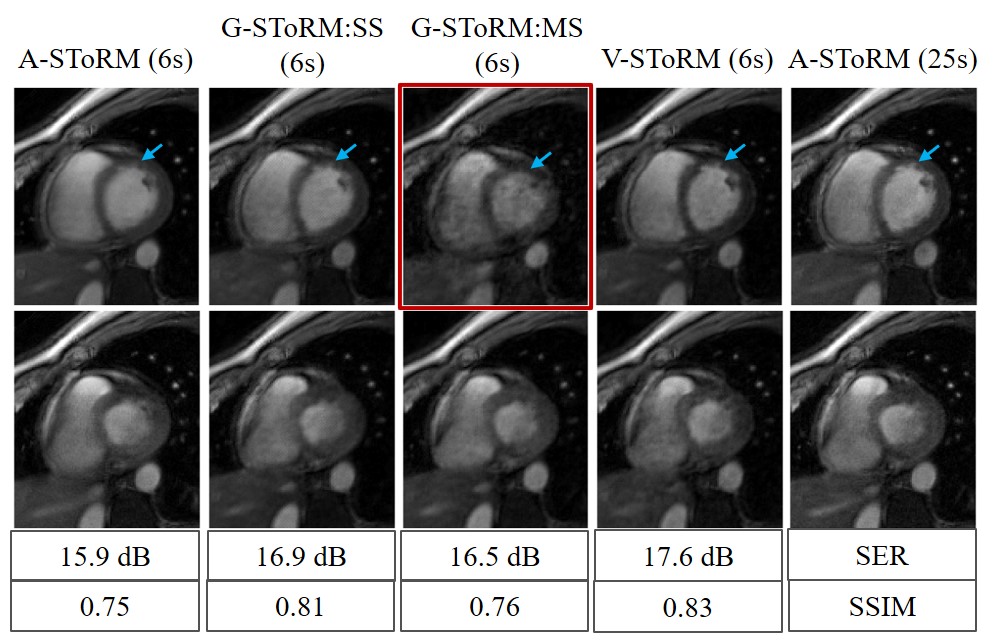}}\\\vspace{-0.7em}
          \subfigure[Latent vectors (V-SToRM)]{\includegraphics[width=0.225\textwidth]{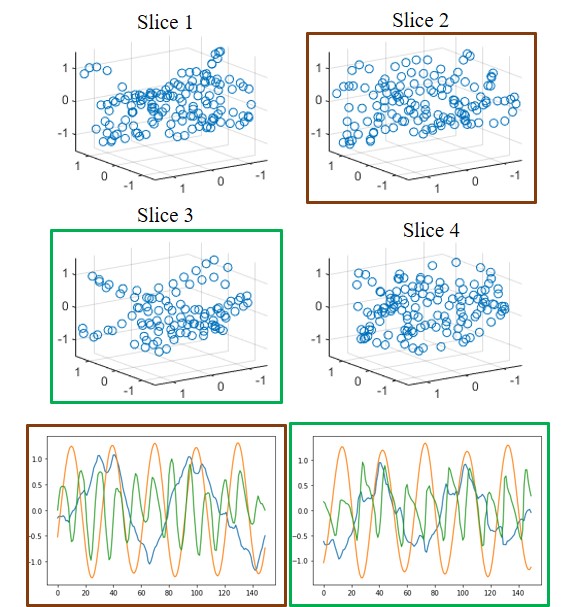}}
          \subfigure[Latent vectors (G-SToRM:MS)]{\includegraphics[width=0.225\textwidth]{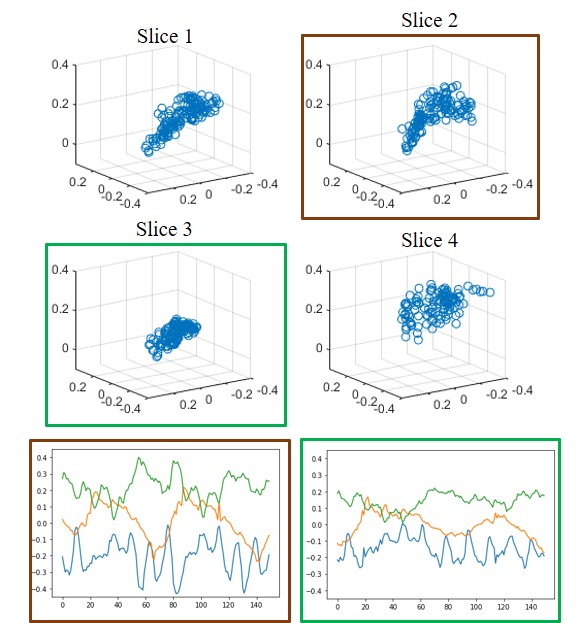}}\vspace{-1em}
	\caption{\small Comparisons with state-of-the-art methods. We show two of the four slices recovered using V-SToRM with two different cardiac phases. The first four columns represent the reconstruction from A-SToRM, G-SToRM:SS, G-SToRM:MS, and the proposed multislice V-SToRM based on the data of 6 seconds.  The distribution of the latent vectors estimated by V-SToRM and multislice G-SToRM from four slices are shown in (c) and (d), together with the time course of latent vectors from slice 2. We note that the priors on the latent vectors in V-SToRM result in the same probability distributions for different slices, while they are very different for G-SToRM. The mismatches in distributions translate to blurred reconstructions with G-SToRM.}\vspace{-1.5em}
	\label{ms_compare}
\end{figure}

In this section, we demonstrate the ability of the algorithm to jointly align and reconstruct free-breathing and ungated multislice dynamic MRI and compare it with state-of-the-art methods. In Fig. \ref{ms_compare}, we compare the image quality of the multislice V-SToRM from six seconds/slice of acquisitions with the image quality of the reconstructions from state-of-the-art manifold methods, including single-slice analysis SToRM (A-SToRM) \cite{ahmed2020free}, single-slice generative SToRM \cite{zou2021dynamic}, and the direct extension of G-SToRM to the multislice setting using a 3D generator. The reconstructions of two slices in a four-slice dataset are shown in the top two rows. For comparisons, we manually picked the matching timeframes reconstructed by different methods. The A-SToRM reconstructions from 25 seconds of data are shown in the last columns of the top two rows, which are used as a reference for quantitative comparisons. We note that A-SToRM was shown to be more competitive than the state-of-the art single-slice approaches in \cite{ahmed2020free}. 

Once the multislice models are jointly trained from the k-t space data of all slices, the latent vectors of the second slice (bottom row) were used to excite the generator. The comparisons show that the V-SToRM reconstructions closely match the performance of 25 s A-SToRM (last column), even though the acquisition time was fourfold lower. The distribution of the latent vectors from first four slices are shown above the latent vector plots.  As a result of the scaling ambiguity, G-SToRM learning latent vectors from different slices have different distributions; exciting the generator with the latent vectors of the second slice results in poor reconstructions. By contrast, the priors within V-SToRM ensure that the latent vectors have the same distribution, which translates to good reconstructions. The comparisons with single-slice methods clearly show the improved signal-to-noise ratio (reported in dB) as well as reduced artifacts; we attribute the improved performance over G-SToRM:SS to the ability of V-SToRM:MS to exploit the inter-slice redundancies. 

\begin{figure}[!htpb]
\centering
	\includegraphics[width=0.3\textwidth]{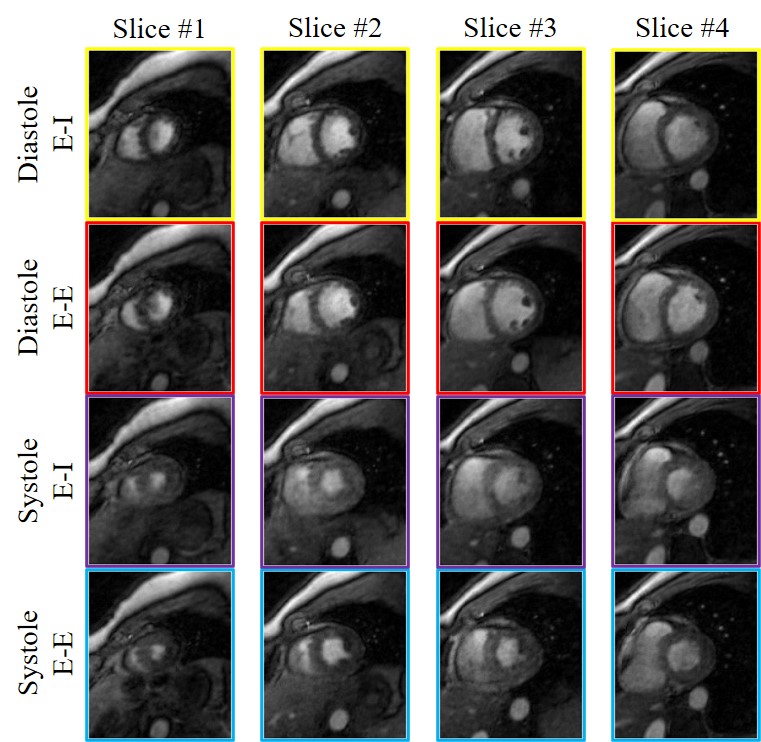}
	\caption{\small Mulislice and multi-phase reconstructions offered by V-SToRM on a dataset with four slices. We fed the latent vectors corresponding to the second slice into the generator. Four different phases in the time series for each slice are displayed. }
	\label{showcase}
\end{figure}

In Fig. \ref{showcase}, we illustrate the ability of the multislice-V-SToRM to offer the alignment of the dataset with four slices. In Fig. \ref{showcase}, we show diastole in end-inspiration, diastole in end-expiration, systole in end-inspiration and systole in end-expiration for each slice. From Fig. \ref{showcase}, we see that the proposed multislice V-SToRM scheme is able to jointly reconstruct and align the multislice free-breathing and ungated cardiac MRI. We note that all the slices in the same dataset have the same cardiac phase and respiratory phase, even though the slices are acquired sequentially with different motion patterns.

\section{Conclusion}

We introduced a  variational framework for the learning of a CNN manifold model from undersampled measurements. This work generalized the traditional VAE scheme to the undersampled setting; unlike the traditional VAE scheme that uses an encoder to learn the conditional distribution from the images, we propose to learn the parameters of the distribution from the measurements using back-propagation. The application of the framework to multislice cardiac MRI data enabled the joint alignment and recovery from highly undersampled measurements. Unlike current single-slice methods that perform the independent recovery of the slices, the proposed approach aligns the acquisitions and jointly recovers the images from the undersampled k-t space data. In addition to facilitating the exploitation of inter-slice redundancies, this approach also eliminates the need for post-processing schemes to match the phases of the slices.

\section{Compliance with Ethical Standards}
\vspace{-1em}
This research study was conducted using human subject data. The Institutional Review Board approved the acquisition of the data, and written consent was obtained from the subjects.

\bibliographystyle{IEEEbib}
\bibliography{refs}
\vspace{-1em}
\end{document}